\documentclass[preprint,12pt]{elsarticle}

\usepackage{amssymb}
\journal{Physics Letters A}

\begin{document}

\begin{frontmatter}


\title{Desynchronizing Networks Using Phase Resetting}
%\tnotetext[label1]{}
\author[a]{Jon Borresen\corref{cor1}}
\cortext[cor1]{Corresponding Author} \ead{J.Borresen@mmu.ac.uk}
\author[b]{David Broomhead}
\address[a]{School of Computing, Mathematics and Digital
Technology, Manchester Metropolitan University, Manchester UK M1
5GD}
\address[b]{School of Mathematics, The University of Manchester, Manchester
UK M60 1QD}

%\title{Desynchronizing Networks Using Phase Resetting}
%
%%% use optional labels to link authors explicitly to addresses:
%\author[a]{Jon Borresen\corref{cor1}}
%\author[b]{David Broomhead}
%\address[a]{School of Computing, Mathematics and Digital
%Technology, Manchester Metropolitan University, Manchester UK M1
%5GD}\cortext[cor1]{Corresponding Author}\ead{J.Borresen@mmu.ac.uk}
%\address[b]{School of Mathematics, The University of Manchester, Manchester
%UK M60 1QD}

\begin{abstract}
\noindent Understanding complex systems which exhibit
desynchronization as an emergent property should have important
implications, particularly in treating neurological disorders and
designing efficient communication networks. Here were demonstrate
how, using a system similar to the pulse coupling used to model
firefly interactions, phase desynchronization can be achieved in
pulse coupled oscillator systems, for a variety of network
architectures, with symmetric and non symmetric internal oscillator
frequencies and with both instantaneous and time delayed coupling.
\end{abstract}


\end{frontmatter}

% \linenumbers

\section{Introduction}
\label{sec:Intro} \noindent Synchronization, both full and partial,
has been observed and studied in a wide variety of systems and is
known to be an essential feature of the collective dynamics of
interacting systems
\cite{Newmann,Pikovsky2,Boccaletti1,Gauthier,Nishimura}.
Desynchronization is likewise of considerable interest. This is the
case particularly in neuroscience where it has implications for the
treatment of Parkinson's disease \cite{Hurtado,Payoux} and other
neurological disorders such as epilepsy \cite{Sowa,Jones,Pinto}.
Consequently, research has focussed primarily on breaking
synchronization in neural systems via a variety of methods
\cite{Tass1,Tass2,Tass3,Coombes1,Bressloff1,Goldobin} or
investigating phase resetting, cluster splitting and the stability
of cluster dynamics in general
\cite{Dolan1,Ashwin1,Kori1,Kori2,Zhai,Shkarayev}. Desynchronization
is also a useful concept in designing asynchronous communication
systems \cite{Olfati,Boyd,Mutazono,Pagliari}.

Arguably,  desynchronization is a ubiquitous naturally occurring
{\em{emergent property}}. Here we demonstrate a general mechanism,
similar to that found in the pulse coupling model of synchronization
in firefly interactions, which generates desynchronization in a
variety of coupled oscillator network architectures, having either
uniform or non-uniform distributions of frequencies and either
instantaneous or time-delayed coupling. We use an adaptation of the
Mirollo-Strogatz model \cite{Mirollo}, which is itself an adaptation
of the Peskin \cite{Peskin} model for synchronization in heart
pacemaker cells. (See also \cite{Timme,Timme2,Wu,Ernst1,Popovych}
for investigations of coupled systems with time delays.)

\section{The Basic Model}
\label{sec:basic}
\begin{figure}[h!]
\begin{center}
\includegraphics*[height=30mm, width=50mm]{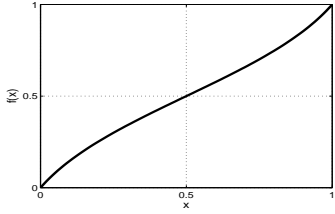}
\end{center}
\caption{\label{fig:IntFun} Interaction function $f$ (Equation
\ref{eqn:IntFun}) has a stable fixed point at $x=0.5$ and an
unstable fixed point at $x = 0$.}
\end{figure}

Consider a network of $N$ coupled oscillators as an
undirected graph, $(V,E)$, where the vertices, $V$, represent
individual oscillators and the edges, $E$, the network connections
between them. The state of each vertex $v_i\in V$, where
$i\in(1,\ldots,N)$ is described by a phase $x_i\in\mathcal{S}^1$. In
this basic model, it will be assumed that the internal oscillator
frequencies are identical and that the coupling is instantaneous.
This model will later be extended to cover variable frequencies and
networks with time delays. In a manner similar to \cite{Mirollo}
coupling is through phase resetting. Whenever the phase of an
individual oscillator passes through $x_i=1$ it is reset to $x_i=0$
($0\sim 1$ in $\mathcal{S}^1$) and each oscillator connected
directly to this oscillator has its phase reset according to a
smooth function $f:\mathcal{S}^1\rightarrow \mathcal{S}^1$ where
$Df(x)>0$ for all $x\in  \mathcal{S}^1$. For desynchronization to
occur, $f$ will be a nonlinear function having the asynchronous
state as a stable (under iteration) fixed point with all other fixed
points unstable. For instance
\begin{equation} \label{eqn:IntFun}
f(x)=\frac{1}{4}\left(\ln(1+(e^2-1)x)-\ln(1+(e^{-2}-1) x)\right),
\end{equation}
(inspired by the rise function of the Mirollo-Strogatz model) has an
unstable fixed point at $0$ (the synchronous state) and a stable
fixed point at $0.5$ (see figure \ref{fig:IntFun}).
%%%%%fig%%%%%%%

%%%%%%%%%%%%

For a network consisting of oscillators with identical frequencies
and instantaneous coupling, steady-state behaviour will be
determined by the network topology and the form of $f$. In the
simplest network, consisting of two coupled oscillators with phases
($x_1,x_2$), the dynamics is a flow on a $2$-torus (seen as the unit
square with the opposing edges identified). Since the oscillators
advance at the same rate, the flow is parallel to the diagonal
$x_1=x_2$. The identification of the opposite edges means that
trajectories leaving the right and top edges of the square reappear
at the left and bottom edges respectively (equivalent to resetting
$x_i=1$ to $x_i=0$). Function evaluations $f(x)$ are performed at
this time. The form of $f$ yields a repelling closed orbit $x_1=x_2$
(as the gradient $f'(1)>1$ ) and an attracting closed orbit
$x_1=x_2-0.5$, since $0<f'(0.5)<1$.

\begin{figure}[h!]
\begin{center}
\includegraphics[angle=0, origin=c, width=45mm, height=30mm]{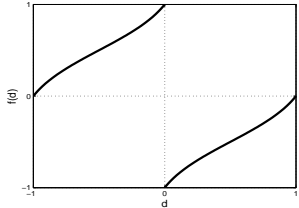}
\end{center}
\caption{\label{fig:Map2} Discrete map function for two oscillator
model describing phase difference $d=x_1-x_2$ in the rotating $x_1$
frame. This represents the second return map for the flow and
function evaluations for each phase rotation of $x_1$.}
\end{figure}

Considering the dynamics in the rotating $x_1$ frame, the
$2-$dimensional system may be reduced to a $1-$dimensional map
describing the phase difference, $d=x_1-x_2$. We observe that the
phase rotation is an isometry which swaps the upper and right edges
of the unit square, where the function evaluations occur. The phase
difference after each rotation of $x_1$ is given by a single
discontinuous function, $F$, which is the composition of the phase
rotation and $f$ (see Figure \ref{fig:Map2}).
\begin{equation} \label{eqn:Diff2}
F(d)=\left\{
\begin{array}{ll}
 f(d+1) & -1 < d < 0 \\
0 & d=0 \\
-f(1-d) & 0 < d \leq 1
\end{array} \right.
\end{equation}
%%%%fig%%%%%%%%

$F(d)$ has a globally attracting period-$2$ orbit $d=\pm 0.5$ (the
desynchronized state $x_1=x_2-0.5$). The other (isolated) fixed
point ($d =0$) is unstable. For the two oscillator model, therefore,
the asymptotic dynamics are a globally stable limit cycle at
$x_1=x_2-0.5$; apart from initial synchronization, all initial
conditions result in phase desynchronization whereby the difference
between the phases of the oscillators is maximal.

For a system of $N$ globally coupled oscillators---with phases
$\mathrm{x}=(x_1,\ldots,x_N)$---the dynamics can be represented as
flow in an $N$-torus, seen as an $N$-cube with opposing faces
identified. The flow is parallel to the $x_1=\ldots=x_N$ diagonal
and function evaluations occur on the exiting faces.  As with the
two oscillator model, we determine the steady state behavior by
considering a discrete map equivalent of the dynamics constructed
from the phase differences in the rotating $x_1$ frame i.e.
$d_{i}=x_1-x_{i+1}$. The transformation from $\mathrm{x}$ to
$\mathrm{d}$ is accomplished by a $(N-1)\times N$-matrix $M =
(\mathrm{e}:-\mathbf{1}_{N-1})$ where $\mathrm{e}$ is a column
vector with all components equal to unity and $\mathbf{1}_{N-1}$ is
the $(N-1)\times (N-1)$ unit matrix. Between transitions each
component of $\mathrm{d}$ is conserved because the flow is parallel
to the diagonal.

If the components of $\mathrm{x}$ are ordered so that $x_1>\ldots
>x_N$, the first transition occurs when $x_1=1$ and the effect of
the functional mapping is $\mathrm{d}\mapsto G(\mathrm{d})$ where
$G_k(\mathrm{d})=-f(1-d_k)$. With all-to-all coupling (since
$x_i>x_{i+1}$ implies  $f(x_i)> f(x_{i+1})$), following the
transition the cyclic permutation $P_{\ast}(x_1, x_2, \ldots, x_N)=
(x_2, x_3, \ldots, x_1)$ restores the assumed ordering of the
components of $\mathrm{x}$. Given a general permutation $P$, we
would like to find a corresponding transformation $T$ on
$\mathrm{d}$, where $T M=M P$. Since $M P$ is a matrix consisting of
the permuted columns of $M$, we write $M P =(\mathrm{v}:M')$ where
$\mathrm{v}$ is the column coming first in the permutation, and $M'$
is the rest of the matrix. Now $T M = T
(\mathrm{e}:-\mathbf{1}_{N-1}) = (T\mathrm{e}:-T)$ which implies
that $T = -M'$. For this to be a solution, it must be true that
$\mathrm{v} = -M'\mathrm{e}$. This follows from the fact that the
row sums of $M$ are all zero. The transformations $\{T\}$ inherit
the group structure of the set of permutation matrices $\{P\}$. In
particular, if $\{P\}=<P_\ast>$, the cyclic group of order $N$
generated by $P_\ast$, then the group of transformations, $\{T\}$,
is isomorphic, consisting of the powers of $T_\ast$ where $T_\ast M
= M P_\ast$. Given any initial condition which is consistent with
the initial ordering of the phases we can evolve $\mathrm{d}$ in
time
\[
\mathrm{d}(t+k) = T^{1-k}_\ast \circ G\circ(T_\ast \circ G)^{k-1}
(\mathrm{d}(t)).
\]
When $k=N$, the order of the cyclic group, this reduces to a simple
iteration
\[
\mathrm{d}(t+N) =(T_\ast \circ G)^{N} (\mathrm{d}(t)).
\]

%%%%fig%%%%%%%%
\begin{figure}[h!]
\begin{center}
\includegraphics[angle=0, origin=c, width=70mm, height=60mm]{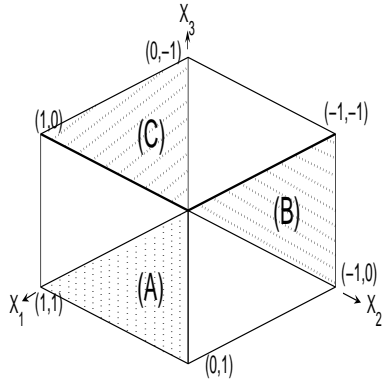}
\end{center}
\caption{\label{fig:Faces} The projection of the 3-cube along its
principle diagonal $x_1=x_2=x_3$. The plane is parameterised by
$\mathrm{d}=(x_1-x_2, x_1-x_3)$ . Points on different faces of the
cube with the same vector $\mathrm{d}$ are identified by the
projection. For example, the shaded region {\bf A}, when interpreted
as being in the bottom face of the cube is the set
$\{x_1>x_2>x_3=0\}$; interpreted as the top face it is the set
$\{1=x_1>x_2>x_3\}$.}
\end{figure}
%%%%%%%%%%%%

The three oscillator all-to-all network is illustrated in Figure
\ref{fig:Faces} which shows a projection of the 3-cube along its
principle diagonal $x_1=x_2=x_3$. In this projection the flow
reduces to a field of fixed points. Points on different faces of the
cube which correspond to the same vector $\mathrm{d}$ are identified
by the projection. The shaded region {\bf A}, when interpreted as
being in the bottom face of the cube is the set $x_1>x_2>x_3=0$. The
flow identifies this with the top face $1=x_1>x_2>x_3$
interpretation of {\bf A}. Resetting $x_1$ to zero and applying $f$
to the other two phases maps this region onto {\bf B} (interpreted
as the back face). The same argument now shows that {\bf B} maps
onto {\bf C} and thus back to {\bf A}. Region {\bf A} is therefore
invariant under the action of  $(T_\ast \circ G)^{3}$. The closure
of {\bf A} contains synchronized or partially synchronized states in
its boundary. However, the choice of $f$ ensures that these are all
unstable. Numerical work suggests that given this form of $f$, the
interior of {\bf A} contains a unique attracting fixed point. That
is, we find a unique attracting orbit which visits the regions {\bf
A}, {\bf B} and {\bf C} cyclically. The unshaded regions contain a
second attracting orbit which is generated by assuming a different
initial ordering of the phases, $x_2>x_1>x_3$ for example. In the
general $N$ clock setting, assuming the unique attractor suggested
by the numerics, the group theoretical underpinning of this system
allows us to count the number of distinct orbits corresponding to
desynchronized states. The order of the symmetric group consisting
of all permutations of $N$ objects is $N!$ and, by Lagrange's
theorem, the number of permutations which are not mapped to each
other when acted upon by $<P_{\ast}>$ is $(N-1)!$. For each of these
initial permutations there is an attracting desynchronized orbit and
a repelling synchronized orbit (excluding symmetry).

\section{Systems with Non-identical Frequencies}
\label{sec:NIF}
\begin{figure}[h!]
\begin{center}
\includegraphics[angle=0, origin=c, width=50mm, height=30mm]{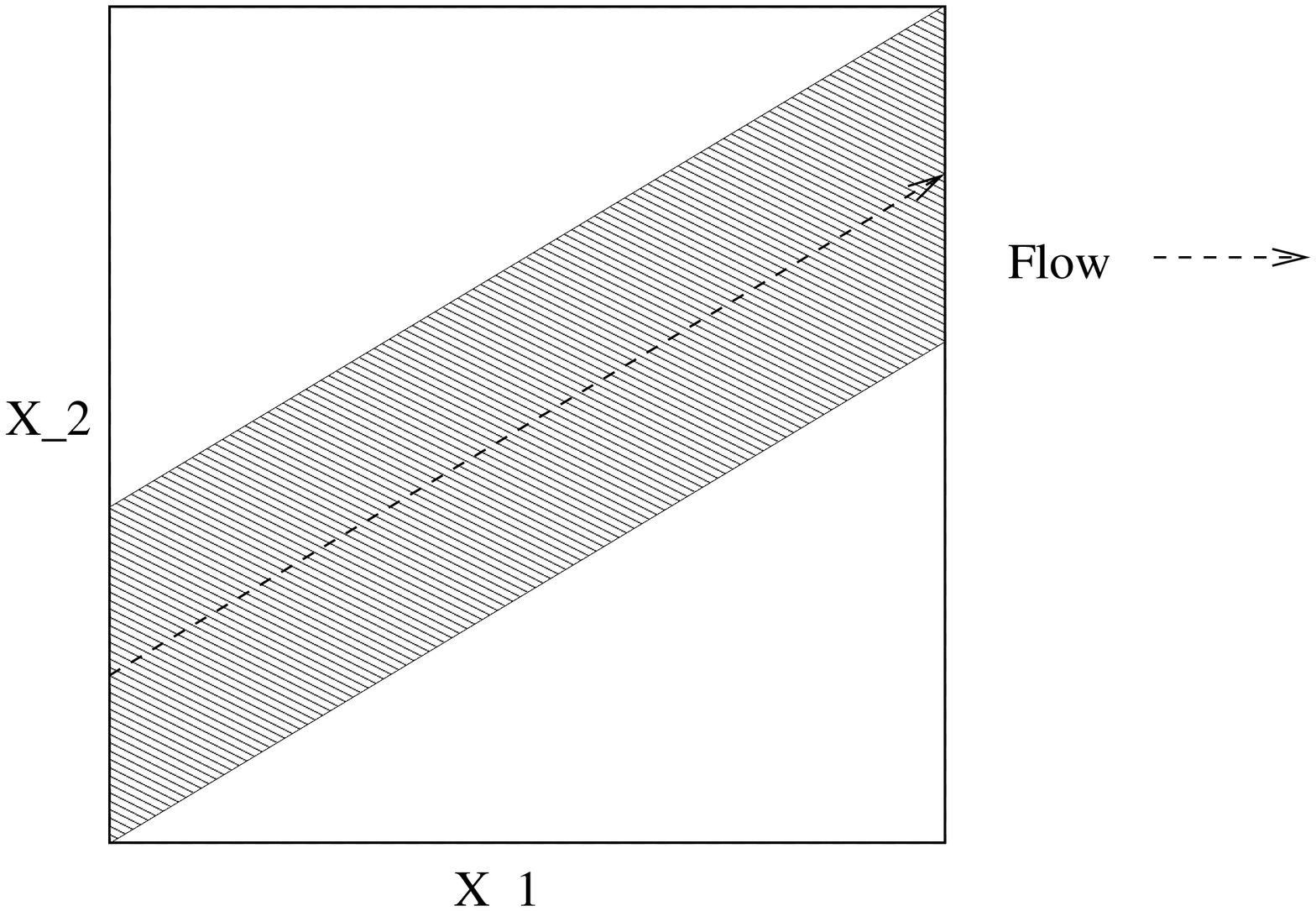}
\end{center}
\caption{\label{fig:Overtake2} When the flow is not parallel to the
diagonal $x_1=x_2$, the oscillators will alternately pass through
$x_i=1$, if the limit cycle is outside the hashed region.  In this
case, desynchronization will occur. The size of the hashed region is
determined by the interaction function $f(x)$ and the angle of flow.
}
\end{figure}

The heart of the previous analysis is that the
behaviour of the $N$-oscillator system is based on a cyclic
permutation of the initial ordering of the phases. We might suppose
that this property is robust to a sufficiently small variation in
the frequencies of the oscillators. Consider a system of two
oscillators with frequencies $\omega_1$ and $\omega_2$.  We can
demonstrate graphically that if $\rho=\frac{\omega_1}{\omega_2}$ is
less than some critical value $\rho_c$ then limit cycles as
described above will occur. Figure \ref{fig:Overtake2} demonstrates
that if a stable limit cycle exists outside a region determined by
the ratio of the oscillators' frequencies then the dynamics will
pass through alternate faces of the torus in succession.  As the
change in flow does not affect the existence of the globally stable
limit cycle (as the interaction function $f$ has not changed), it is
only required that the limit cycle lie within this region.
%%%%%fig%%%%%%%
%%%%%%%%%%%%
It is straightforward to derive the orbit of the limit cycle to be
$x_i=f(\rho(1-f(\rho(1-x_i))))$ for all $x_i$.  For $f$ given in
equation \ref{eqn:IntFun} the limit cycle is within the
`non-overtaking' region (see figure \ref{fig:Overtake2}) for $\rho
\leq 1.11$. This value increases as the gradient of equation
\ref{eqn:IntFun} becomes more pronounced.

Using a similar argument, it can be shown that for any network it is
sufficient that the ratio of any two connected oscillators'
frequencies be less than $\rho_c$ for desynchronization to occur.
Again, the value of $\rho_c$ is determined by the interaction
function $f$. For networks which are not globally connected, we can,
via an argument identical to that used above, demonstrate that any
two connected oscillators will desynchronize. However some
consideration must be given to the reducibility of such networks via
symmetry arguments (see \cite{Stewart}). For non-globally connected
networks, phase coupling of the form proposed will result in
{\em{local asymptotic desynchronization}} across the network i.e.
each oscillator will desynchronize with each of those to which it is
connected.

\section{Time Delays}
\label{sec:TD}
The final modeling assumption that will be relaxed
concerns time delays across the network.  When considering
desynchronization using the previous coupling function (equation
\ref{eqn:IntFun}) the time delay can reverse the stability of the
fixed point at $x=0$. Consider two oscillators separated by a time
delay of $\tau$.  If the difference in their phases $x_1-x_2<\tau$,
when the first oscillator sends a pulse on crossing $x=1$, the
second oscillator passes $x=1$ before receiving the pulse.  In this
case the second oscillator is perturbed closer to the first and with
each cycle the oscillators move closer together.

\begin{figure}[h!]
\begin{center}
\includegraphics[angle=0, origin=c, width=30mm, height=30mm]{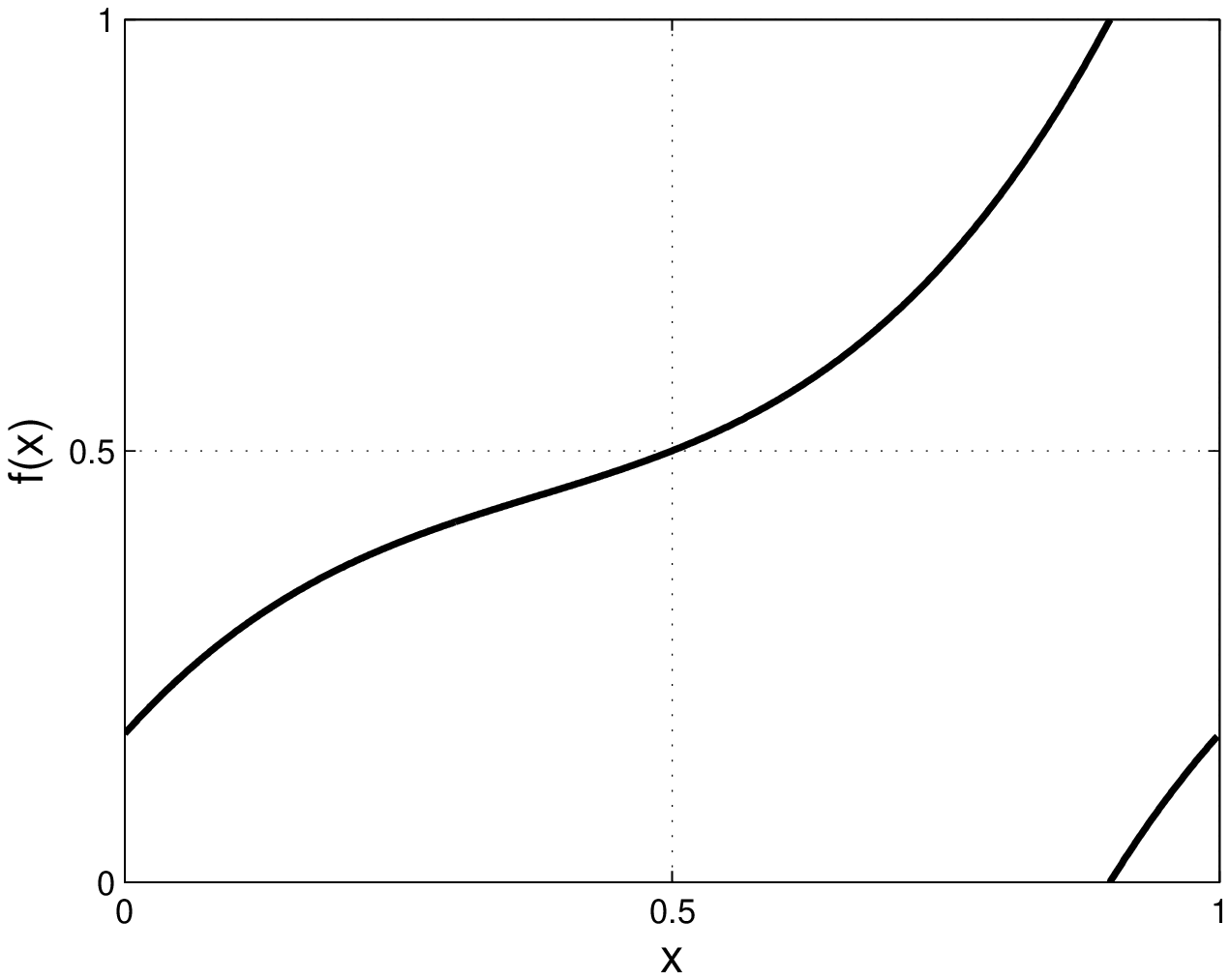}~~
\includegraphics[angle=0, origin=c, width=30mm, height=30mm]{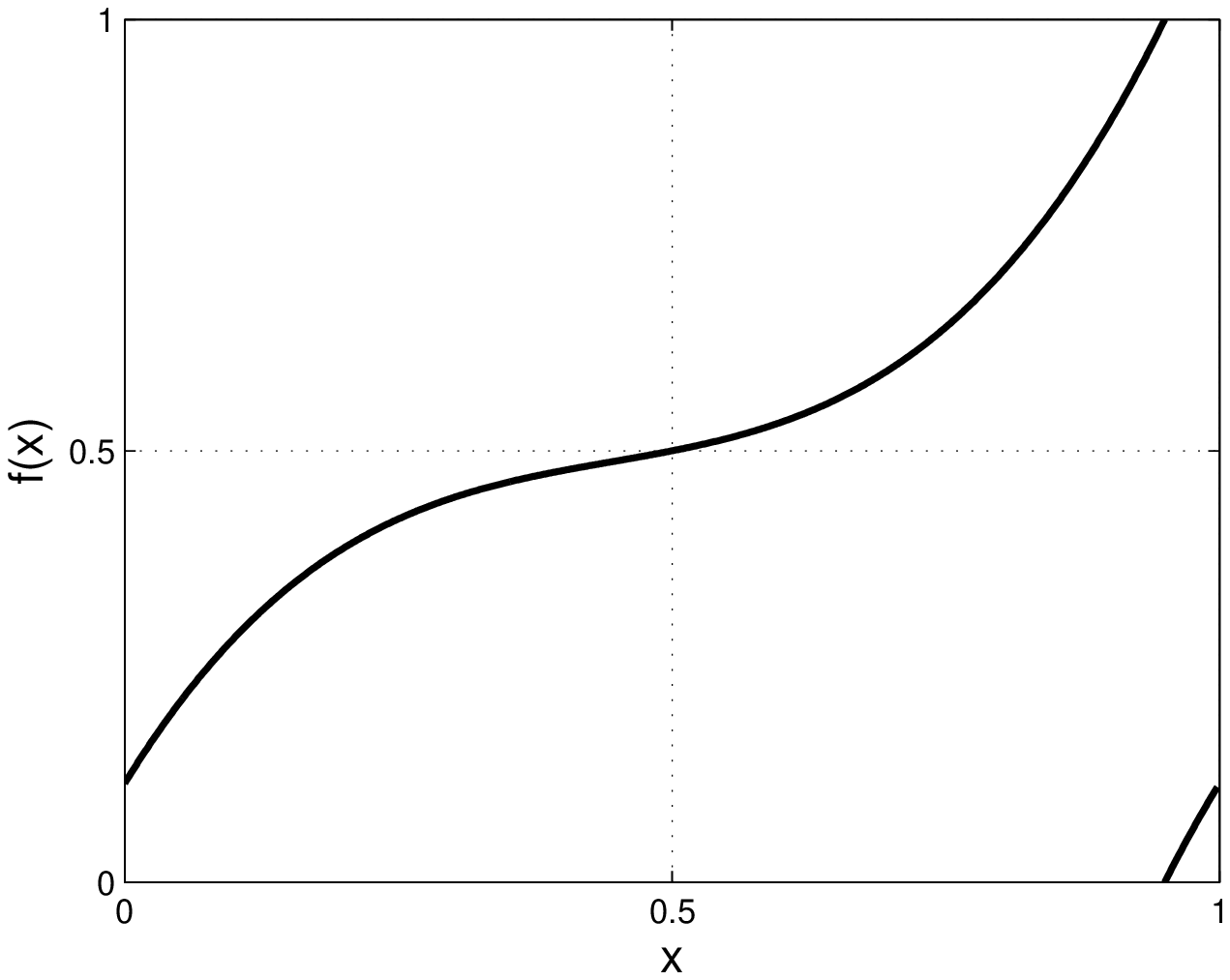}
\end{center}
\caption{\label{fig:Delay} Discontinuous interaction functions with
a stable fixed point at $x=0.5$ and no fixed point at $x=0,1$. Left
figure $\tau=0.1$ and $\beta=0.5$ and right $\tau=0.05$ and
$\beta=0.25$.}
\end{figure}

This difficulty can be overcome by redesigning the interaction
function $f$. From the argument described above synchronization will
only occur (using a continuous interaction function) if the
oscillators' frequencies differ by less than the propagation delay
between them. It is required, therefore, that should this occur, the
perturbed oscillator (the one receiving the data) should not be
perturbed closer to the transmitting oscillator's time. We can
derive a new interaction function which has this property and still
retains a stable fixed point at $d=0.5$ using a shifted cubic curve.
Interpolating through the fixed point at $d=0.5$, the curve can be
expressed as follows (see also Figure \ref{fig:Delay}) if $\tau$ is
the maximum propagation delay, and $\beta\in (0,1)$ the gradient at
the fixed point:
\begin{equation}
f(x)=a(\tau,\beta)x^3+b(\tau,\beta)x^2+c(\tau,\beta)x-\tau \,
\mathrm{mod} \, 1.
\end{equation}

This function is a continuous mapping of the circle
$f:S^1\rightarrow S^1$ but appears discontinuous in the interval $x
\in [0,1)$.
%%%%%fig%%%%%%%
%%%%%%%%%%%%

There are important factors within the transient dynamics, which may
have some impact on the functioning of such time delayed systems.
For instance, the existence of transient chaos cannot, at this
stage, be excluded nor other dynamical effects present in systems of
interacting oscillators and particularly the presence of unstable
attractors within the dynamics needs to be considered
\cite{Ashwin2}. However, the above arguments can be applied to
suggest that the only asymptotically stable dynamics of such systems
would be the desynchronized state, where each oscillator pulses in
turn.
%%%%%fig%%%%%%%
\begin{figure}[h!]
\begin{center}
\includegraphics[angle=0, origin=c, width=90mm,
height=50mm]{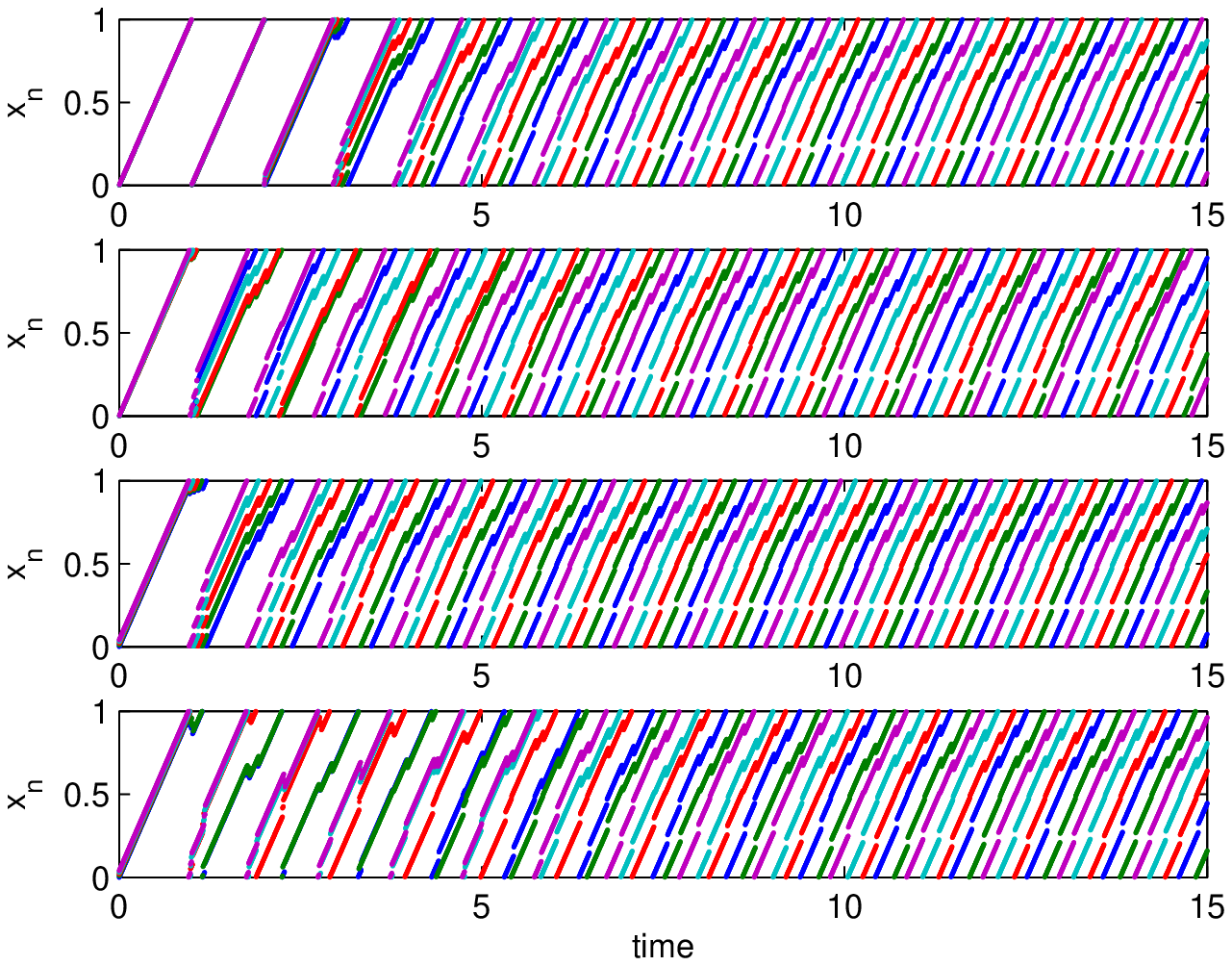}
\includegraphics[angle=0, origin=c, width=90mm,
height=50mm]{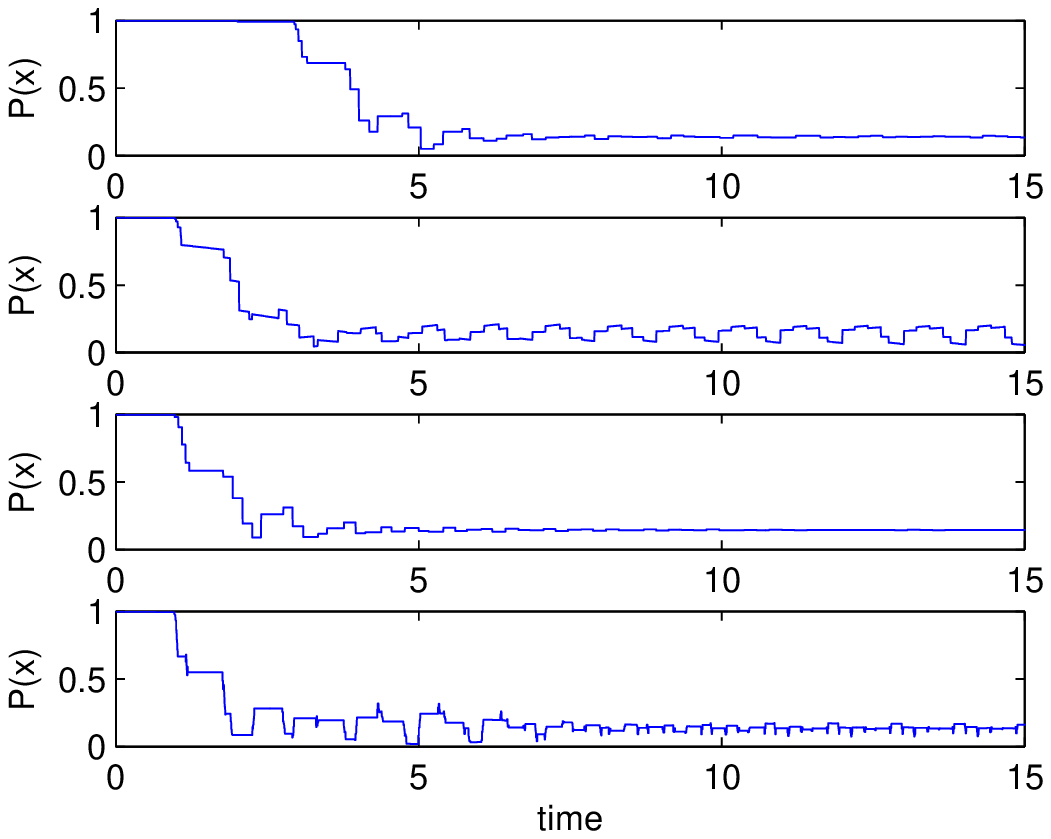}
\end{center}
\caption{\label{fig:Sims} Time series for
 simulations of $5$ globally coupled oscillators (top) and the
 associated time series for the order parameter $P$ (bottom) with near synchronous initial
 conditions: (a) identical
oscillator frequencies period $=1$, no time delays; (b) normally
distributed frequencies (mean period $=1$ standard deviation
$=0.05$); (c) identical oscillator frequencies, uniform time delay
$\tau=0.01$; (d) normally distributed frequencies (mean period $=1$
standard deviation $=0.05$) uniform time delay $\tau=0.01$.}
\end{figure}
%%%%%%%%%%%%

\section{Simulations}
\label{sec:sims} The model has been simulated for a variety of
networks, both homogeneous and inhomogeneous, using identical
internal frequencies, distributed internal frequencies and for
networks with small, uniform time delays.  In all cases local
asymptotic desynchronization was observed, in accordance with the
above (see Figure \ref{fig:Sims}).  For time delayed systems the
duration of transient behaviour grows considerably as the number of
oscillators is increased and it is possible to observe clustering if
a suitable choice of the interaction function is not made, however,
the long term dynamics appear in all cases to converge to the
desynchronized limit cycle previously described.\\

Figure \ref{fig:Sims} also shows the time series for an order
parameter $P$ which gives a measure of the total coherence in the
network \cite{Hansel, Daido}.
\begin{equation}
P=\frac{1}{N}| \sum_{k=1}^{N}e^{2 \pi i x_k}|,
\end{equation}
where $P=1$ corresponds to synchronous oscillation and for any other
state $0\leq P <1$.  As can be observed, with initial conditions
near the ordered synchronous state, the oscillator dynamics rapidly
desynchronize and the order of each oscillator as it is distributed
around the phase remains unchanged.\\

\section{Summary}
\label{sec:sum} The analysis as presented here demonstrates how, via
pulse coupling, a network of connected oscillators may be forced to
achieve phase desynchronization as a collective dynamic. The model
is intended to demonstrate `proof of principle' of the design of an
emergent property.  We conjecture that the method of pulse coupling
applied here would be equally applicable to weakly coupled
oscillators exhibiting synchronization. The applications of such a
concept may be far reaching, particularly when applied to digital
communication systems, the design of neural based computers and in
the treatment of Parkinson's disease and epilepsy.
%%%%%%%%%%%%
{\em{Acknowledgements}} The authors would like to thank J. Shapiro,
M. Sorea, S. Furber and L.O.Gowrie for their advice, discussions and
encouragement. This work was sponsored by EPSRC grant GR/T11258/01.


%% References


\bibliographystyle{model1a-num-names}
\bibliography{PLABib}






% \end{thebibliography}

\end{document}